\documentclass[journal]{new-aiaa}

\usepackage{graphicx,subcaption}
\usepackage{epstopdf,color}
\usepackage{cleveref}
\usepackage{xspace,doi}
\usepackage{longtable,tabularx}

\def\fullline{\hbox to\textwidth}

\def\beq{\begin{equation}}
\def\eeq#1{\ifx\string#1\string \nonumber\else\label{#1}\fi\end{equation}}

\newcommand\revised[1]{\textcolor{blue}{#1}}

\makeatletter
\def\mylabel#1{ \stepcounter{figure}%
 \write\@auxout{%
    \string\newlabel{#1}{{\arabic{figure}}{\thepage}{}{figure.\arabic{figure}}{}} %
  }}
 \makeatother

\def\newbracketequ#1#2\eeq#3{\beq\left.
  \parbox{#1\textwidth}{\vskip-3em\begin{eqnarray*}#2\nonumber\end{eqnarray*}\vskip-2em } 
   \right\} \eeq{#3}
}

\long\def\comment#1\endcomment{}\def\endcomment{}
\def\komeg{\mbox{$k-\omega$}\xspace}

\def\beq{\begin{equation}}
\def\eeq#1{\ifx\string#1\string \nonumber\else\label{#1}\fi\end{equation}}

\def\newequpar#1\eeq#2{\beq \begin{split}#1\end{split}  \eeq{#2}}

\def\mystrut#1{\rule[#1em]{0pt}{.1em}} 

\newdimen\labbx\labbx=1.9em
\newdimen\eqbx\eqbx=\textwidth\advance\eqbx -2em
\def\newequ#1\eeq#2{\hfill\break\parbox{\eqbx}
 {\begin{align*}#1\end{align*}}\hfill%
  \ifx\string#2\string \else
 \parbox{\labbx}{\hfil\begin{eqnarray}\label{#2}\end{eqnarray}}\fi\break
}

\def\mathfrac#1#2{\leavevmode\kern-.2em\hbox{\kern.25em\raise.5ex\hbox{$\scriptstyle #1$}
    \kern-.2em{\small/}\kern.1em\lower.3ex\hbox{$\scriptstyle #2$}}\kern.2em }

\begin{document} 
\title{Modification of the $k-\omega_0$ model for roughness}

\author{ Paul Durbin\footnote{Professor, Department of Aerospace Engineering, and Senior member AIAA, e-mail: durbin@iastate.edu,}}
\affil{Iowa State University, Ames, Iowa 50014}
\author{Zifei Yin\footnote{Associate Professor, School of Aeronautics and Astronautics, 
Shanghai 200240 China, e-mail: yinzifei@sjtu.edu.cn}}
\affil{ Shanghai Jiao Tong University, Shanghai China}

\date{\vspace{-3ex}}

\maketitle

\begin{abstract}

 Surface roughness plays a substantial role in many flows for which Reynolds averaged prediction is
needed.  The transformation used in the $k-\omega_0$ model is extended to rough surfaces by adding an effective origin.
The  log-layer offset is computed as a function of  this effective origin, thereby creating a correspondence between
effective origin and equivalent sandgrain roughness.  

A formula is derived for the virtual origin of the fully rough
log law.  It is shown how the present model is consistent with the fully rough limit.

\end{abstract}

\section*{Nomenclature}

{\renewcommand\arraystretch{1.0}
\noindent\begin{longtable*}{@{}l @{\quad=\quad} l@{}}
$A$  & constant in  \cref{eq:Transform} \\
$B$ & additive constant in log-law\\
$C_{\omega1,2}$  & coefficients of the $\omega$ equation, standard values 5/9, 3/40 are used\\
$C_{\mu}$  & coefficient of the $k$ equation \\
$\ell$ & effective origin\\
$r$ & equivalent sandgrain roughness\\
$|S|$ & magnitude of the rate of strain\\
$\Delta U$ & rough wall log-layer offset\\
$y_{0}$ & virtual origin of the log-lay\\
$\nu$  & molecular viscosity \\
$\omega_{0}$ &   regular solution to $\omega$ equation \\
$\omega$ &   standard, singular solution to $\omega$ equation \\
$\tau_{w}$ & wall shear stress
\end{longtable*}}

\section{Introduction}

Surface roughness plays a substantial role in many flows for which Reynolds averaged prediction is
needed \citep{Jimenez}.  Even surfaces that are nearly smooth geometrically, may be hydrodynamically rough at 
high Reynolds number. If $r$ is the roughness height, 
the flow is affected when $r_{+}\equiv ru_{*}/\nu=R_\tau\, r/\delta $ is not small.  If $R_\tau$ is a few thousand, and
$\delta$ is a centimeter, roughness can be significant when $r$ is on the order of 
a few 10's of microns.

Because the viscous wall layer is very thin in high Reynolds number turbulence, it can be disturbed 
by turbulent eddies created by small surface irregularity
\citep{RoughnessReview}.  The mechanism by which these eddies 
are created is not well understood. The
dominant eddies are larger than the scale of the roughness \citep{Umair}.  They are generated on local shear layers, not directly by
the roughness elements. 
\revised{ Whatever the generation mechanism, for the purposes of RANS modeling, their only
property is that they disrupt the  wall layer, which alters the mean flow.}

An operational approach is adopted in RANS modeling:  
the surface is characterized by an equivalent sandgrain roughness height.  This roughness height can be quite
different from the geometrical roughness height.  It is treated as a hydrodynamic property of the surface.
It is determined from a data correlation.  That correlation was measured long ago by Nikuradse,
and correlates an offset of the mean velocity in the log-layer to sand grain size. A convenient curve fit to the correlation was provided by 
\citet{Ligrani}, and recited in \cref{L-M}.   
\revised{Similar roughness correlations are reviewed in \citet[][\S9.2]{RoughnessReview}}.

With an eddy viscosity model, the log-layer velocity offset is
determined by the distribution of eddy viscosity between the wall and the log-layer.  
Thus, roughness enters as a modification to the turbulence model within that region: 
such modification is the approach adopted herein.

The purpose of the present paper is to adapt the $k-\omega_0$ model \citep{k0paper} to rough surfaces.  Methods
to adapt \komeg-type models for surface roughness began with \citet{Wilcox}, who introduced roughness via a boundary
value of $\omega$; he left the $k$ boundary condition as $k=0$.  A $k$ boundary condition that interpolates
between zero and a fully rough value was introduced in \cite{Roughke}, along with a treatment of roughness by
adding an \emph{effective origin} to model formulas.  \citet{Knopp} developed a rough-wall \komeg model using similar methods.  
These and other approaches to modeling roughness are reviewed in \citet{RoughJOT}.  The present
approach is to incorporate roughness via an effective  origin.

\section{Log-layer offset}

Surface roughness is characterized by the \emph{equivalent sandgrain roughness}, $r_+$. The rough-wall log-law is
\beq
U_+=\frac1\kappa\log(y_+)+B +\Delta U_+(r_+).
\eeq{log-law}
$\Delta U_+$ is the \emph{velocity offset,} that is cited in the introduction.
$\Delta U_+(0)=0$, and $\Delta U_+(r_+)$ becomes negative as $r_+$ increases.
The standard $k-\omega$ model gives $0.408$ for the VonKarman constant, $\kappa$. 
$B$ is an additive constant that appears in the log-law for a smooth wall.

  The $r_+$ of a surface is found from a calculation of $\Delta U_+$ and a 
  correlation  between it and $r_+$.  The formulas
\newbracketequ{.75} 
\Delta U_+= 0 \hskip12em& r_+<r_{smooth}\\
\Delta U_+=  \xi\,\left(8.5  - B - \log(r_+)/\kappa\right) \qquad&  90>r_+>r_{smooth}\\
         \Delta U_+= 8.5 - B - \log(r_+)/\kappa \hskip3.5em& r_+>90\\
\hbox to0em{\hskip-2em with} \qquad\xi = \sin\left[\frac\pi2 
  \log(\mathfrac{r_+}{r_{smooth}}) / \log(\mathfrac{90}{r_{smooth}}) \right]%
  \mystrut{1.7} 
\eeq{L-M}
 were suggested by Ligrani and Moffat as a fit to the Nikuradse data.
$r_{smooth}$ is the solution to
\beq
8.5  - B - \log(r_{smooth})/\kappa=0
\eeq{rsmooth} 
If  \cref{L-M} is abbreviated as $\Delta U=R(r_+)$, then a calculation of $\Delta U$ determines the equivalent
sandgrain roughness via $r_+=R^{-1}(\Delta U)$.
\revised{ In the transitional range, $r_{smooth}<r_{+} <90$, the correlation can depend on the particular
rough geometry;   \cref{rsmooth} is for uniform, sand grains  \citep{RoughnessReview}.}

Turbulence models can be calibrated in the same way  \citep{Roughke}.
First, introduce an effective origin $\ell$ into the model.  Then,
solve in a fully developed channel flow.  Evaluate the log-layer displacement as 
average of \cref{log-law} over the log-region, $80<y_{+}<0.2\delta_{+}$:
\beq
B +\Delta U_+= \frac1{0.2H_+-80}\int_{80}^{0.2H_+}  U_+(y_+|\ell)-\frac1\kappa \log(y_+)\ dy_+.
\eeq{DeltaU}
(In channel flow $H_+\equiv R_\tau$).  $B$ is evaluated with $\ell=0$.  
The calculated $\Delta U(\ell)$ and \cref{L-M} provide the calibration $\ell_+(r_+)$:
given a surface, characterized by its equivalent sandgrain
roughness, the effective origin, $\ell$, of the model is found from this calibration.

\section{The $k-\omega_0$ equations}

The $k-\omega_0$ model is intended to avoid the requirement of computing a singular solution to the $\omega$ equation.  It preserves the limiting behavior $\omega\propto y^{-2},\ y\to0$ by
transforming $\omega_0$ into $\omega$ \citep{k0paper} .
The only alteration to  the $\omega$ equation is to replace $\omega$ by $\omega_0$, except in the eddy viscosity:
\begin{align}
\frac{D\omega_0}{Dt}   &=2C_{\omega1} |S|^{2}-C_{\omega2} \omega_0^{2}+\nabla[(\nu+\nu_{T}\sigma_{\omega})\nabla \omega_0]
\nonumber\\
\frac{D k}{Dt}\   &=2 \nu_{T}|S|^2-C_{\mu} k\omega+\nabla[(\nu+\nu_{T}\sigma_{k})\nabla k] 
\label{eq:omeg}\\
\nu_{T}\ &=\frac{ k}\omega\nonumber
\end{align}
The transformation 
 \beq
 \omega=\frac{6\nu}{C_{\omega2}y^2}F(X)
 =\omega_0+\frac{6\nu}{C_{\omega2} y^2}e^{-AX} ,\ \hbox{ where }\ X\equiv\frac{\omega_0 y^{2}}\nu
 \eeq{eq:Transform}
 maps $\omega_0$  to $\omega$.
The value $A=0.012$ gives close agreement between them.  One of two boundary conditions is applied:
either $\omega_0(0)=0$, or $\omega_0= \sqrt{C_{\omega1}/C_{\omega2}} \,|\tau_w|/\mu$.  Both are considered
here.  

\revised{On the smooth wall, the boundary condition is that $k$,
 and hence, $\nu_T$ is zero. But, under rough wall conditions, $k$ and $\nu_{T}$, 
 do not vanish at the wall. } Then the wall stress is obtained as $\tau_{w}=\rho(\nu + \nu_{T}(0))dU/dy(0)$.

The original impetus for the  $k-\omega_0$ model was to improve numerical robustness, without altering predictions
of the \komeg model \citep{Chedevergne,k0paper}.  However, the $\omega$ and $\omega_0$ models are not identical; while they give virtually
the same results in channel flow and boundary layers, predictions of incompressible, separated flow
have been found to differ slightly.  
\revised{In high Mach number boundary
layers, $\omega_0$ is considerably more accurate, because the near-wall 
behavior of $\omega$ is follows compressible scaling \citep{k0paper}.  }

\section{The $\omega_0=0$ boundary condition}

The roughness parameter is introduced as an effective origin, $\ell$, in the transformation (\ref{eq:Transform}):
\newequpar
       X   &= \omega_0 (y+\ell)^2/\nu \\
   \omega&= \omega_0+\frac{6\nu}{C_{\omega2} (y+\ell)^2}e^{-AX} 
\eeq{l-model}
Over `fully rough' walls, the log-layer extends to the effective surface.
To that end,  in addition to \cref{l-model}, the $k$-boundary condition
\beq
k(0) = \min\left(1,\frac{\ell^2_+}{3.9^2}  \right)  \frac{|\tau_w|}{\rho\sqrt{C_\mu}}
\eeq{Krough}
is imposed  \citep{Roughke, Knopp}.  It is based on the wall being fully rough when $r_+=90$; but, 
for convenience of calibration, this is estimated as $\ell_+=3.9$.

\begin{figure}[ht]
\includegraphics[trim= .1in 0in .4in .4in,clip,width=0.5\textwidth]{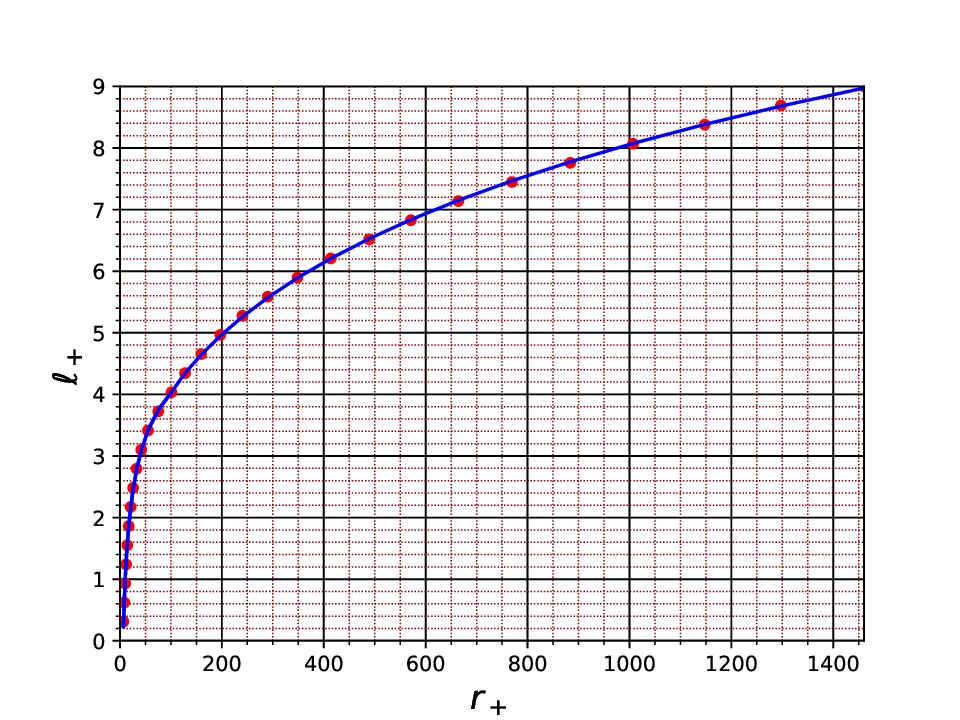}\hskip-3em\raise2em\hbox to 0em{\includegraphics[width=0.35\textwidth]{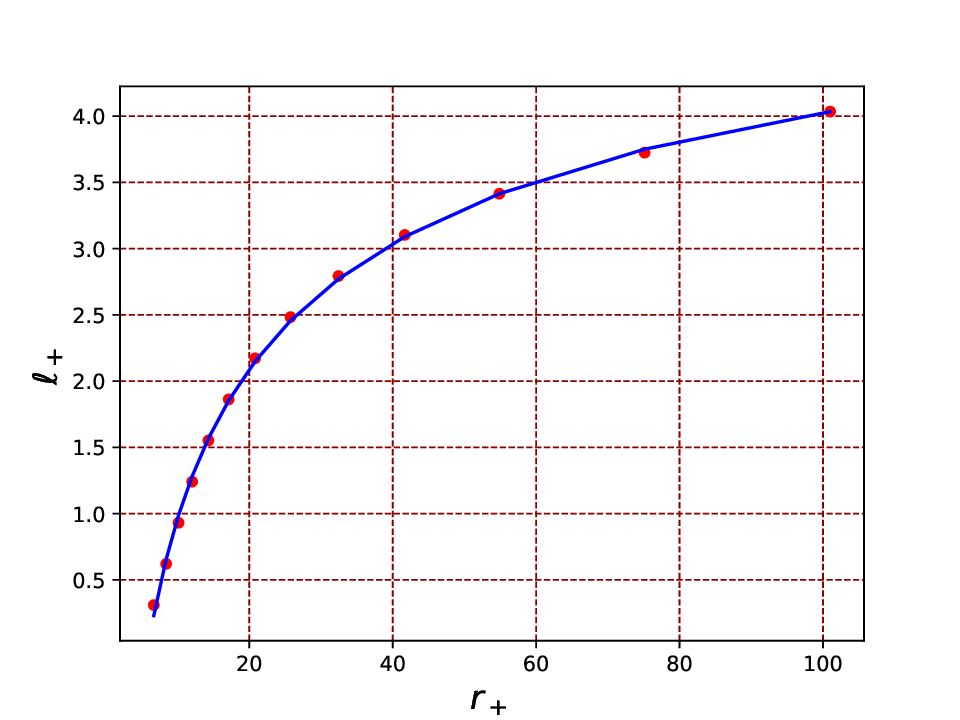}} 				\vspace{-1em}
\caption{Calibration curve, $\omega_0=0$ boundary condition}	 				\vspace{-1em}
\label{fig:calib}
\end{figure}
The model is calibrated by solving it in channel flow at $R_{\tau}=20,000$, then 
finding $\Delta U(\ell_+)$ from \cref{DeltaU}.
Substituting into \cref{L-M} and solving for $r_+$, gives the relation $\ell_+(r_+)$. 
For a smooth wall ($\ell=r=0$)  \cref{DeltaU} gives $B=5.122$.
With this $B$,  $r_{smooth}=3.97$ in (\ref{rsmooth}).

\Cref{fig:calib} is an $\ell_+(r_+)$ calibration curve.  The red circles are computed points. 
The blue curve is fit to those points.
Because $r_+$ increases, about exponentially, with $\ell_+$, the fit is to a polynomial in $\ln r_+$.  
It is split into one polynomial in the transitionally rough  regime, and a second in the fully rough regime:
\newequpar
\ell_+ &= 0,\ r_+< 3.97 \ =\ r_{smooth}\\
\ell_+ &= -0.186(\ln r_+)^2 + 2.614\,\ln r_+ -4.071, \quad  3.97 <r_+<101\\
\ell_+ &= \ \ 0.251(\ln r_+)^2 -1.153\,\ln r_+ + 4.023, \quad   r_+>101.
\eeq{}

\comment
 \begin{figure}[ht!]    
\centering
\includegraphics[angle=0,width=0.5\textwidth]{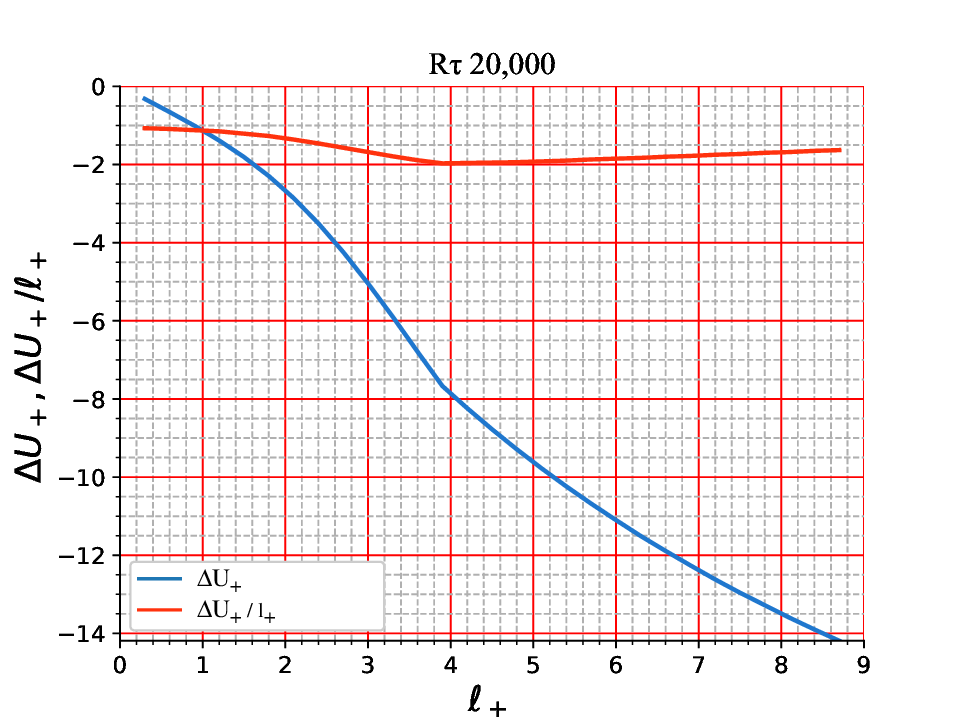}      \vspace{-1em}
\caption{Test whether $\Delta U\propto\ell$}            \vspace{-1em}
\label{fig:DelU-l}
\end{figure}

The ratio $\Delta U_+/\ell_+$ is somewhat constant:
\Cref{fig:DelU-l} shows that $\Delta U_+/\ell_+$ decreases with $\ell$ when transitional rough,
 $k_+<90$, and increases when fully rough, $k_+>90$.  
 This ratio is  $\Delta U_+/\ell_+=-1.63$ at $\ell_+=8.7$
 ($r_+=1,318$).
 It is $-1.70$ at $\ell_+=7.8$  ($r_+=909$).
 It is $-1.96$ at $\ell_+=4.2$  ($r_+=116$).  In fully rough flow, a value of about $-1.8$ might
 be acceptable.
 \endcomment
 
\begin{figure}[ht!]  
\comment
\hbox{\qquad
\vbox to -0.05\textheight{
\hbox{\small Present \qquad\   sandgrain}
\hbox{\small\quad model\qquad\quad roughness}
$\ell_+= 1.2\ \to\ k_+=   11.8$\\
$\ell_+=  2.4\ \to\ k_+=    24.3$\\
$\ell_+=  3.6 \ \to\ k_+=    65.8$\\
$\ell_+=  4.8 \ \to\ k_+=    176.1$\\
$\ell_+=  6.0 \ \to\ k_+=    368.9$\\
$\ell_+=  7.2 \ \to\ k_+=    686.2$\\
$\ell_+=  8.4 \ \to\ k_+=    1157.1$\\ \\
$\ell_+=  3.9\ \to\ k_+=    90$ (fully rough)
}
}
\hskip.45\textwidth
\endcomment
\begin{subfigure}{0.45\textwidth}
\includegraphics[width=\textwidth]{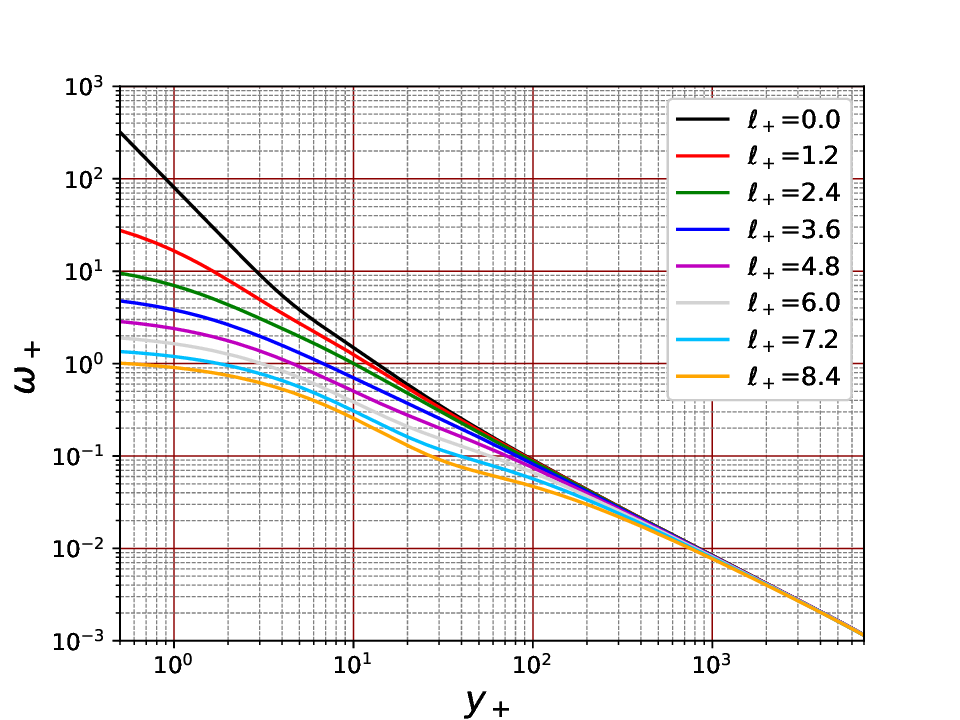}
\subcaption{$\omega$ profiles on smooth and rough walls.}
\end{subfigure}
\begin{subfigure}{0.4\textwidth}
  \includegraphics[trim= 0in 0in .5in .5in,clip,width=\textwidth]{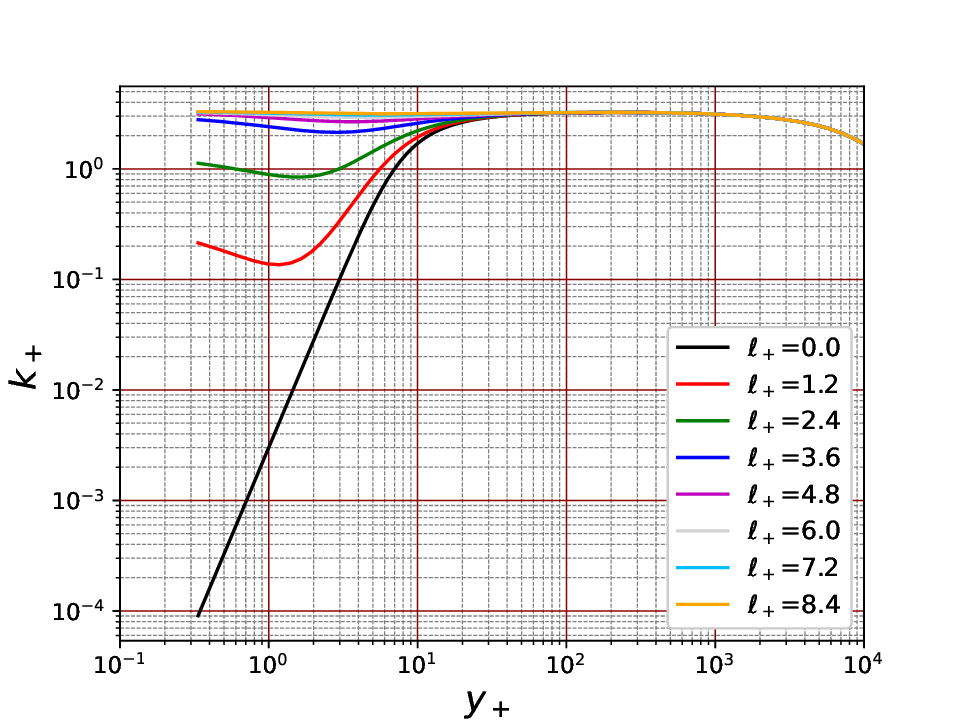}
\subcaption{$k$ profiles. on log-log scale}
\end{subfigure}
\\
\begin{subfigure}{0.4\textwidth}
  \includegraphics[trim= 0in 0in .4in .5in,clip,width=\textwidth]{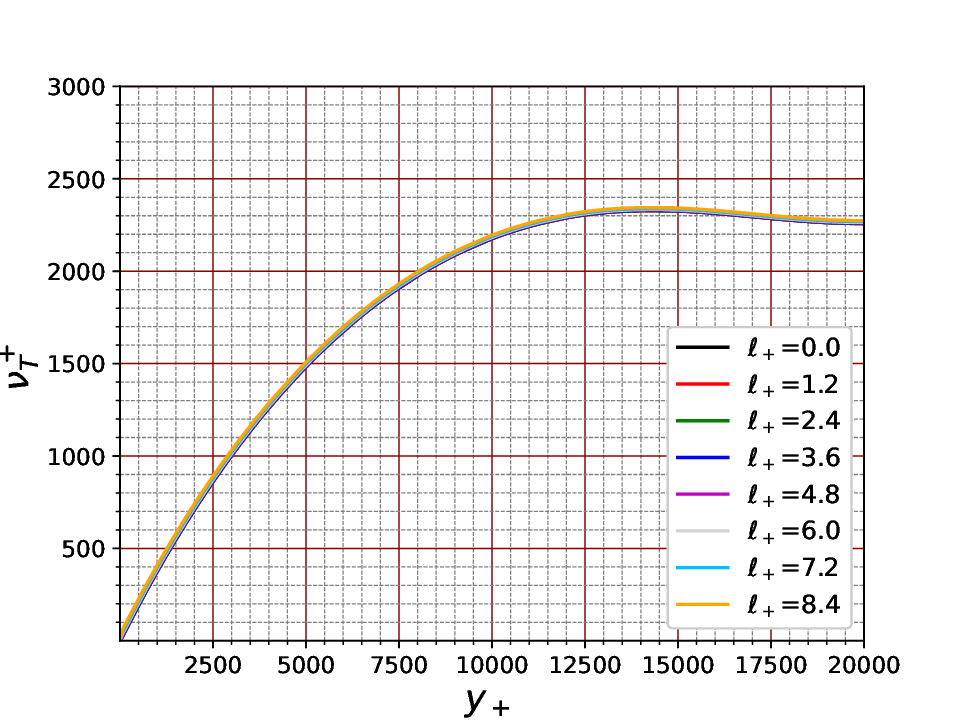}
\subcaption{Eddy viscosity profiles.}
\end{subfigure}
\begin{subfigure}{0.4\textwidth}
  \includegraphics[trim= 0in 0in .5in .5in,clip,width=\textwidth]{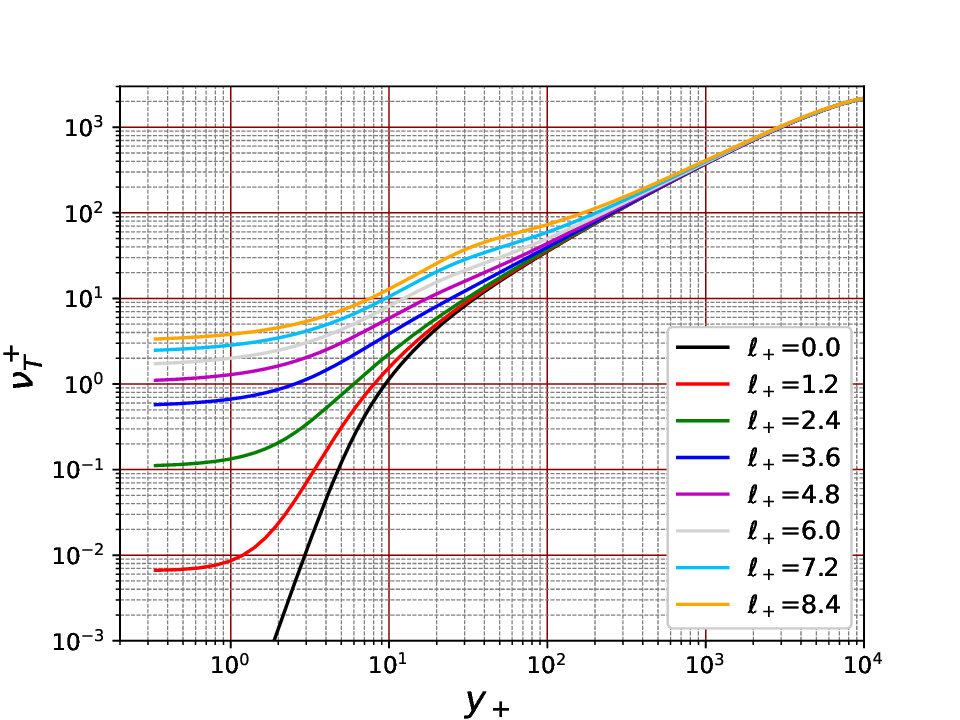}
\subcaption{Eddy viscosity profiles. on log-log scale}
\end{subfigure}\\
\begin{subfigure}{0.4\textwidth}
  \includegraphics[trim= 0in 0in .5in .5in,clip,width=\textwidth]{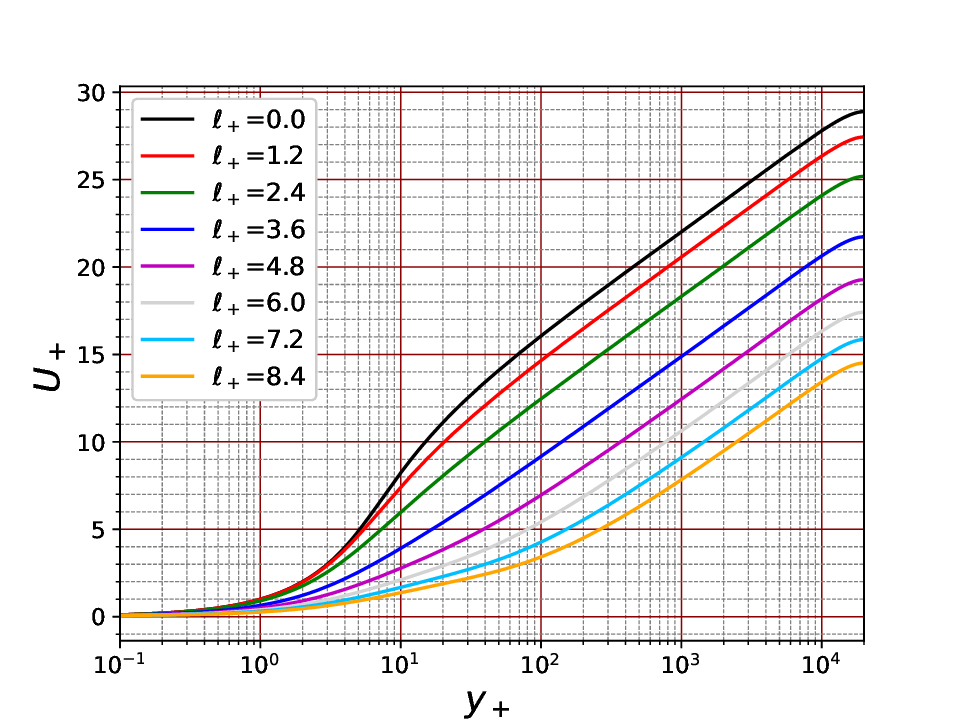}
\subcaption{$U_+$ profiles }
\end{subfigure}
\caption{Effects of roughness via the model.}
\label{fig:BC1}
\end{figure}

 Rough wall solutions are illustrated in \cref{fig:BC1}.  The behavior of $\omega_+$ near the wall is shown in \cref{fig:BC1}a. 
 For a smooth wall $\omega$ is singular like  $1/y^2$, as $y\to0$.  Roughness decreases $\omega$
 dramatically, relieving the smooth-wall singularity.  

 The $k_+$-profiles in \cref{fig:BC1}b show how it transitions between $k_{+}=0$ at a smooth wall, to a constant value
 of 3.3, near fully rough walls.
 
 The eddy viscosity $k/\omega$ is shown in  \cref{fig:BC1}c,d. On linear coordinates, roughness seems to have
 negligible effect; but, on the log-log plot there is, indeed, an effect near the wall.  The velocity gradient varies like 
 $1/\nu_{T}$ so low values of $\nu_{T}$ 
 have a disproportionate influence on the velocity profile.  The behavior near the wall alters
 the whole $U$-profile, as seen in \cref{fig:BC1}e.
 \Cref{fig:BC1}e shows how the log-layer is displaced downward with increasing roughness.  
  At the fully rough condition, $\ell_+=  3.9$, the log-layer extends almost to the wall. 
    
 These figures were computed in channel flow at $R_{\tau}=20,000$.  The largest value of $r_+$ shown 
 is 1,157.  While high, it is small compared to the channel height. 
 
\section{ Shear stress, $\omega_0\propto\tau_w/\mu$, boundary condition}

The other boundary condition, $\omega_0=|\tau_w|/\mu \sqrt{C_{\omega1}/C_{\omega2}}$,
is derived from the source term in the $\omega_{0}$ equation (\ref{eq:omeg})
\citep{k0paper}.  For this boundary condition
the roughness parameter is introduced differently:
\newequpar
       X   &= \omega_0  y^2/\nu \\
  \omega&= \omega_0+\frac{6\nu}{C_{\omega2} (y+\ell)^2}e^{-AX} 
\eeq{ltau-model}
$\ell$ is not included in  the definition of $X$, so that $X$ continues to vanish at $y=0$.
 
\begin{figure}[ht]			
\includegraphics[trim= .1in 0in .4in .4in,clip,width=0.5\textwidth]{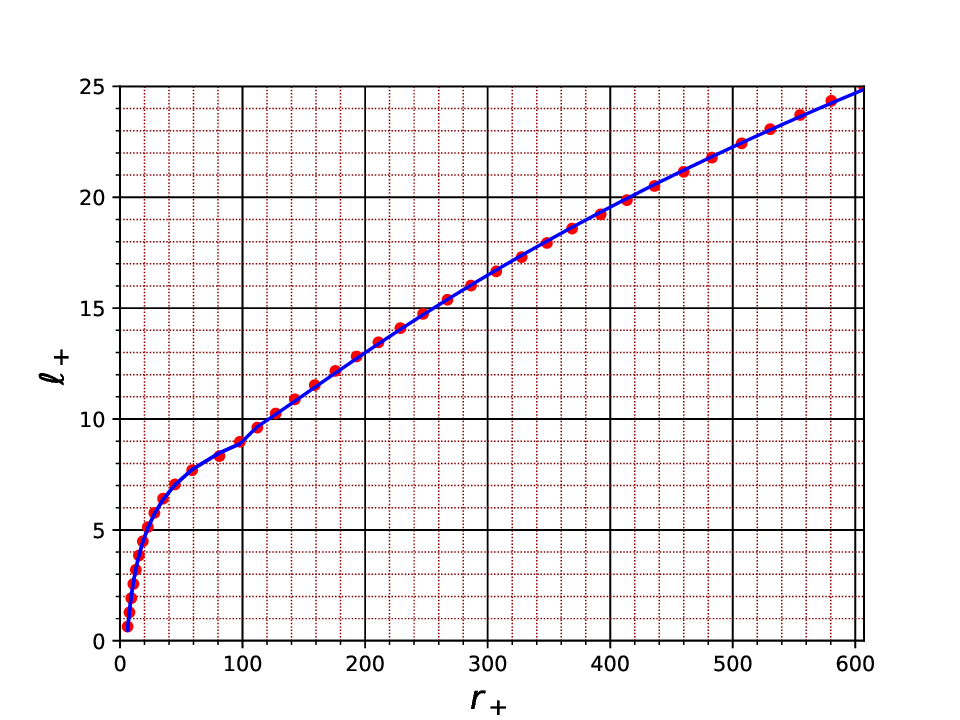}
\hskip-3em\raise2em\hbox to 0em{
\includegraphics[trim= .40in 0in 0in 0in,clip,width=0.33\textwidth]{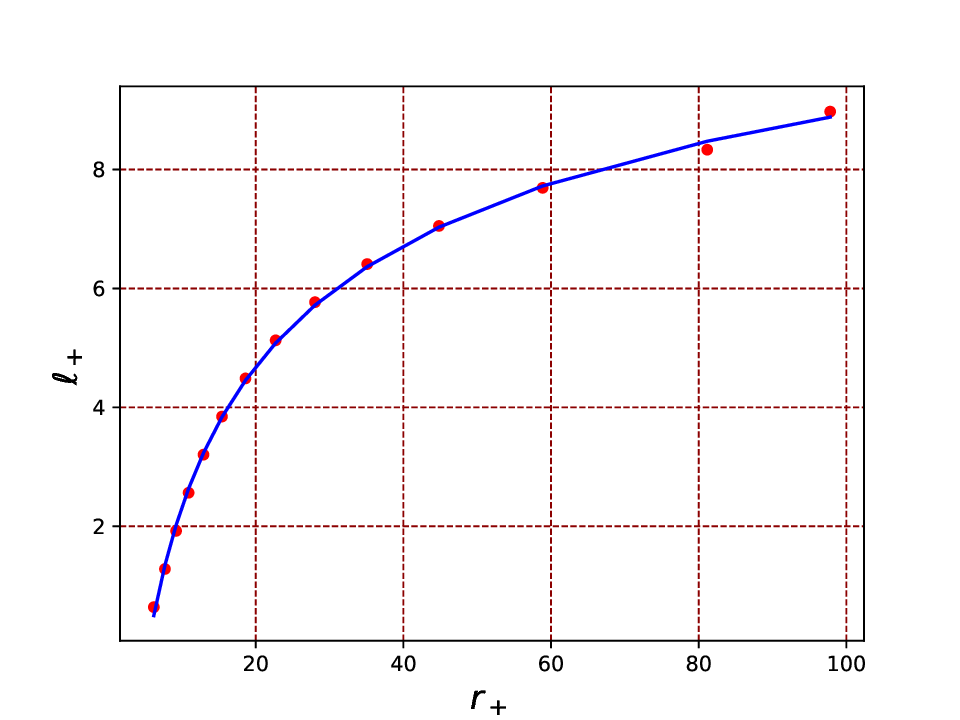}} 				\vspace{-1em}
\caption{Calibration curve, $\omega_0\propto\tau_w/\mu$ boundary condition}	 				\vspace{-1em}
\label{fig:calibltau}
\end{figure}

The model is calibrated as previously, to give \cref{fig:calibltau}, in which the blue curve is
\newequpar
\ell_+ &= 0,\ r_+< 3.76 \ =\ r_{smooth}\\
\ell_+ &= -0.337(\ln r_+)^2 + 5.198\,\ln r_+ -7.861, \quad  3.76 <r_+<97.8\\
\ell_+ &= \ \ 2.924(\ln r_+)^2 -23.540\,\ln r_+ + 55.614, \quad   r_+>97.8
\eeq{}
In this case, the fully rough condition $r_+=90$ is estimated as $\ell_+=8.90$.  The boundary condition
$$
k(0) = \min\left(1,\frac{\ell^2_+}{8.9^2}  \right)  \frac{|\tau_w|}{\sqrt{C_\mu}}
$$
 replaces \cref{Krough}.   
The additive constant on a smooth wall was calculated to be $B=5.252$.  Again, the calibration is at $R_{\tau}=20,000$ (a very similar calibration was found at other, high $R_{\tau}$).

\begin{figure}[ht]			
\comment
\hbox{\qquad
\vbox to -0.02\textheight{
\hbox{\small Present \qquad\   sandgrain}
\hbox{\small\quad model\qquad\quad roughness}
$\ell_+=     2.56\  \ \to\ k_+=  11.4$\\
$\ell_+=   5.12\ \ \to\ k_+=  23.6$\\
$\ell_+=   7.68\ \ \to\ k_+=   61.4$\\
$\ell_+=  10.24 \ \to\ k_+=  134.0$\\
$\ell_+=  12.80 \ \to\ k_+= 203.6$\\
$\ell_+=  15.36 \ \to\ k_+=  281.8$\\
$\ell_+=  17.92 \ \to\ k_+=  367.4$\\
$\ell_+=  20.48 \ \to\ k_+=  459.0$\\
$\ell_+=  23.04 \ \to\ k_+=  559.5$\\
$\ell_+=  8.9\hskip1.3em \to\ k_+=    90$ (fully rough)
}
}
\hskip.44\textwidth
\endcomment
\begin{subfigure}{0.45\textwidth}
  \includegraphics[trim= 0in 0in .47in .47in,clip,width=\textwidth]{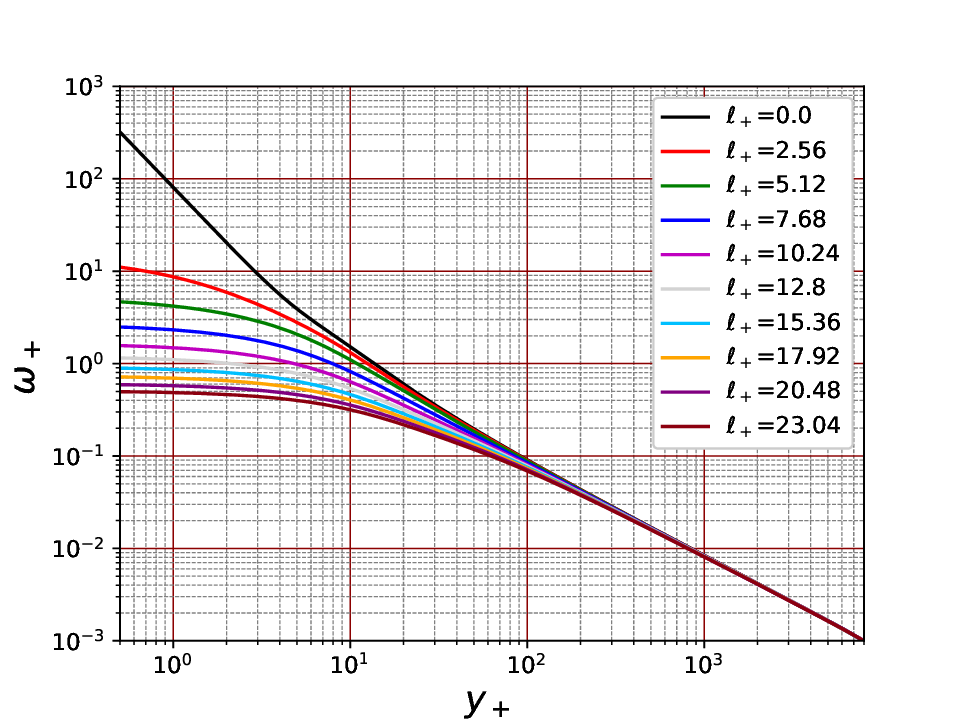}
\subcaption{$\omega$ }
\end{subfigure}
\begin{subfigure}{0.45\textwidth}
  \includegraphics[trim= 0in 0in .47in .47in,clip,width=\textwidth]{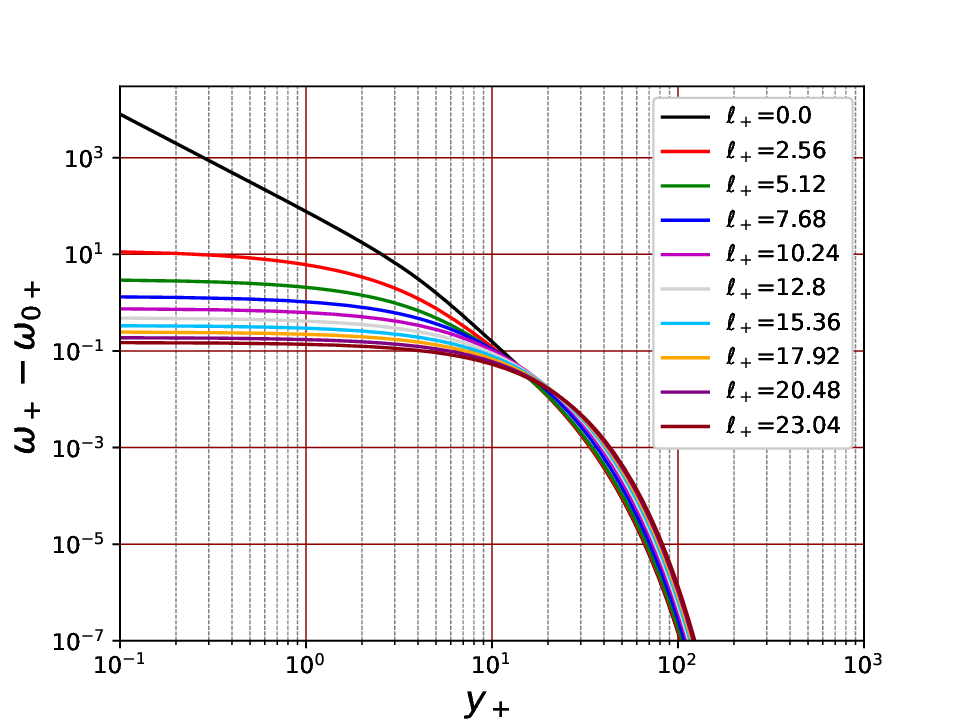}
\subcaption{$\omega-\omega_0$}
\end{subfigure}
\\
\begin{subfigure}{0.45\textwidth}
  \includegraphics[trim= 0in 0in .47in .47in,clip,width=\textwidth]{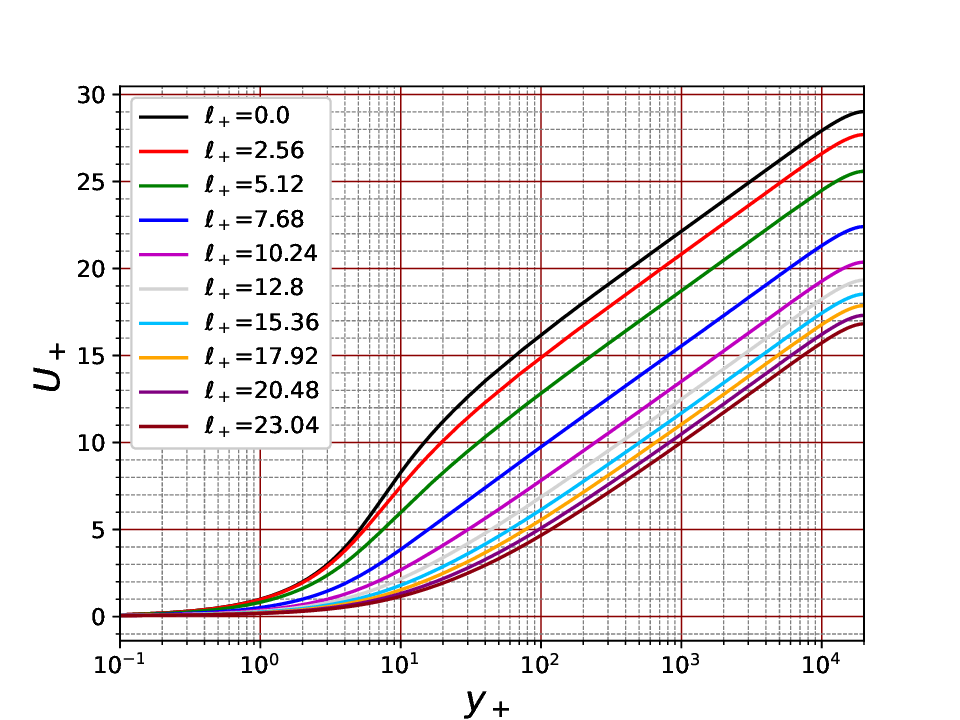}
\subcaption{$U_+$ }
\end{subfigure}
\caption{Effects of roughness with the  $\omega_0\propto\tau_w/\mu$ boundary condition.}
\label{fig:BC2}
\end{figure}

Solutions are quite similar to those with the first boundary condition.  \Cref{fig:BC2}(b) is a plot of
$\omega_{0}-\omega$.  Where this difference is not small, the transformation  (\ref{ltau-model}) is active.
Where it is small $\omega_{0}\approx\omega$, and there is no direct effect of roughness.
The affected region is mainly below $y_{+}=20$.

\section{ Fully rough limit}

The idealized fully rough condition is that the log-layer extends to the wall, with a virtual origin inside the wall.  What
is the virtual origin for the present model?

In the log-layer, $k$ is constant. The $k$-equation (\ref{eq:omeg}) and constant total stress give 
$k_+=1/\sqrt{C_\mu}$.  Also, the $k$-equation reduces to a balance between production and dissipation,
giving $\omega_+=k_+dU_+/dy_+$.  If $\omega_0=\omega$ in the log-layer, which is usually the case
for \cref{eq:Transform}, then the $\omega$-equation
has the solution
\beq
\omega_+=\frac1{\kappa\sqrt{C_\mu}(y_++y_{0+})}
\eeq{LLomega}
$y=-y_{0+}$ is the virtual origin.  Correspondingly, $U_{+}=\mathfrac1\kappa \log(y_++y_{0+})+B+\Delta U$.

If it is assumed that this log-layer solution is valid at a fully rough wall,
equating \cref{LLomega} to \cref{l-model}, with both evaluated at $y=0$, gives the virtual origin
\beq
\frac1{\kappa\sqrt{C_\mu}y_{0+}}=\omega_{0+}(0)+\frac{6}{C_{\omega2} \ell_+^2}\ \to\ 
 y_{0+}=\frac{C_{\omega2} \ell_+^2}{\kappa\sqrt{C_\mu}(6+\omega_{0+}(0)C_{\omega2} \ell_+^2)}
\eeq{offset}
 of the log-law.  
 
 For the second $\omega_{0+}$ boundary condition
 $$
  y_{0+}(r_+)=\frac{C_{\omega2} \ell_+^2}{\kappa\sqrt{C_\mu}(6+\sqrt{C_{\omega1}C_{\omega2}}\,\ell_+^2)}
  =\frac{3\ell_+^2}{\kappa(72+\sqrt{6}\,\ell_+^2)}
 $$ 
where $C_{\omega1} = 5/9$, $C_{\omega2} = 3/40$, $C_{\mu} = 9/100$.
   
\begin{figure}[ht]
\begin{subfigure}{0.45\textwidth}
\includegraphics[trim= 0in 0in .5in .5in,clip,width=\textwidth]{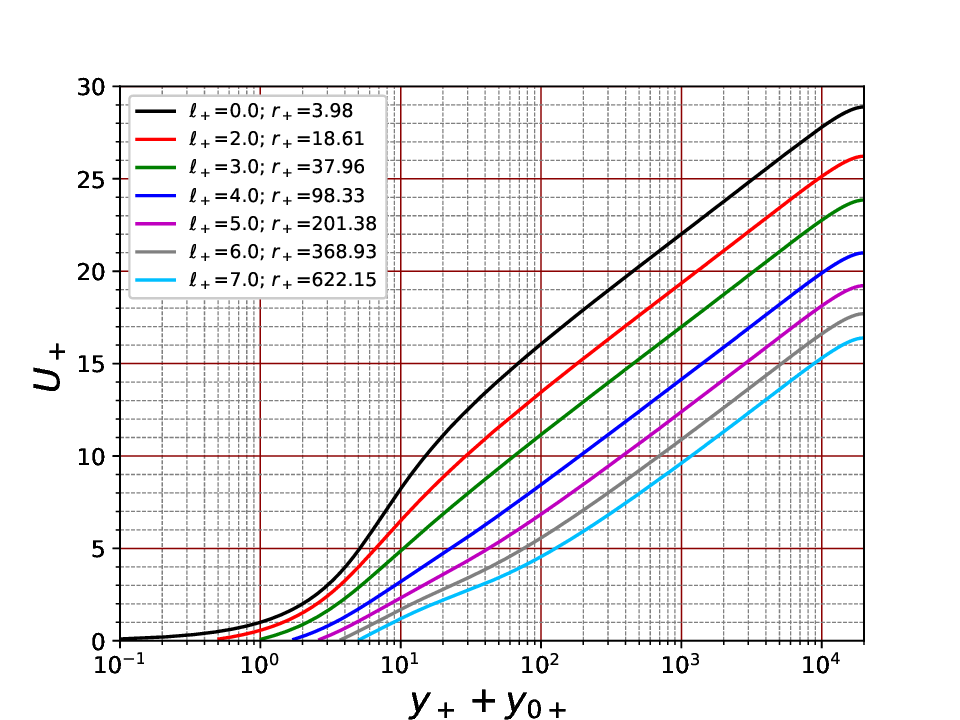}
\subcaption{$\omega_0(0)=0$}
\label{fig:Ushift}
\end{subfigure}
\begin{subfigure}{0.45\textwidth}
\includegraphics[trim= 0in 0in .5in .5in,clip,width=\textwidth]{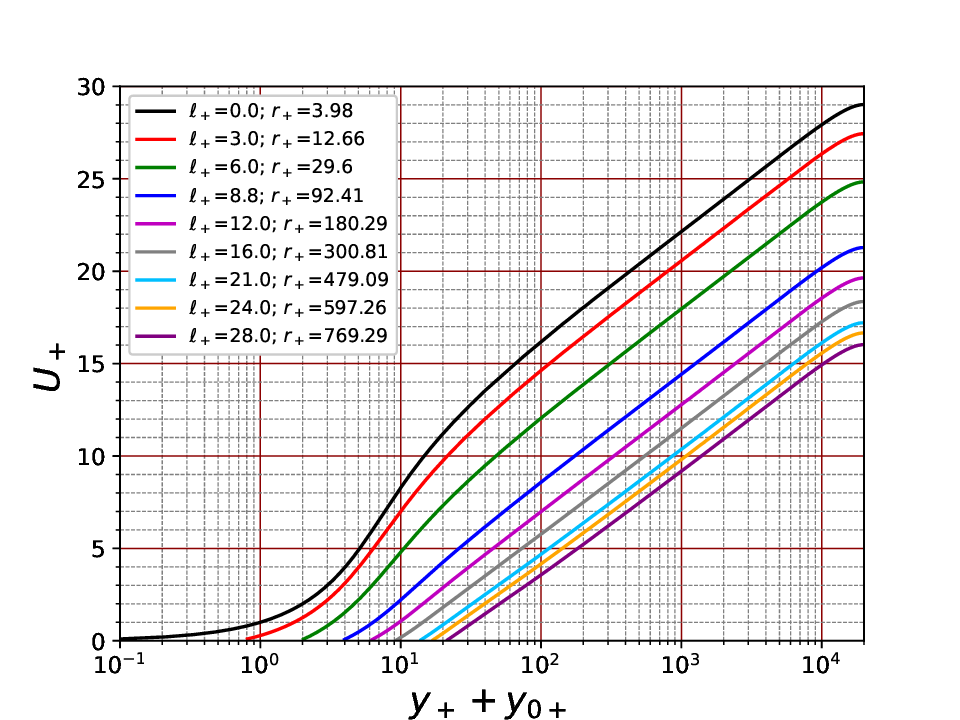}
\subcaption{Shear stress boundary condition}
\label{fig:Ushift1}
\end{subfigure}
\caption{Profiles using virtual origin, $y_+ + y_{0+}$.}
\end{figure}

The  boundary condition $\omega_0(0)=0$ causes
 $\omega_0$ to depart from the log region behavior, and \cref{offset} is not exact.  
 As \Cref{fig:Ushift} shows, in the fully rough cases
 ($\ell_+ > 3.9$), the log-layer extends approximately to the wall, with 
 the estimate \cref{offset} of the virtual origin; but, the curves are not exactly straight
 lines for small values of $y_{+}+y_{0+}$.
 
For the shear-stress boundary condition,
 the second term in \cref{ltau-model} becomes small compared to the first, when $\ell$ increases into the
 fully rough regime.  Then the log-layer can extend to the wall,
and the virtual origin \cref{offset} becomes exact.   \Cref{fig:Ushift1} shows that this is the case: when
 $\ell_+ > 8.90$ the curves are straight lines from the wall, through the log-layer.

\section{Comparisons}

\begin{figure}[ht]
\centering
\begin{subfigure}{0.48\textwidth}
\includegraphics[width=\textwidth]{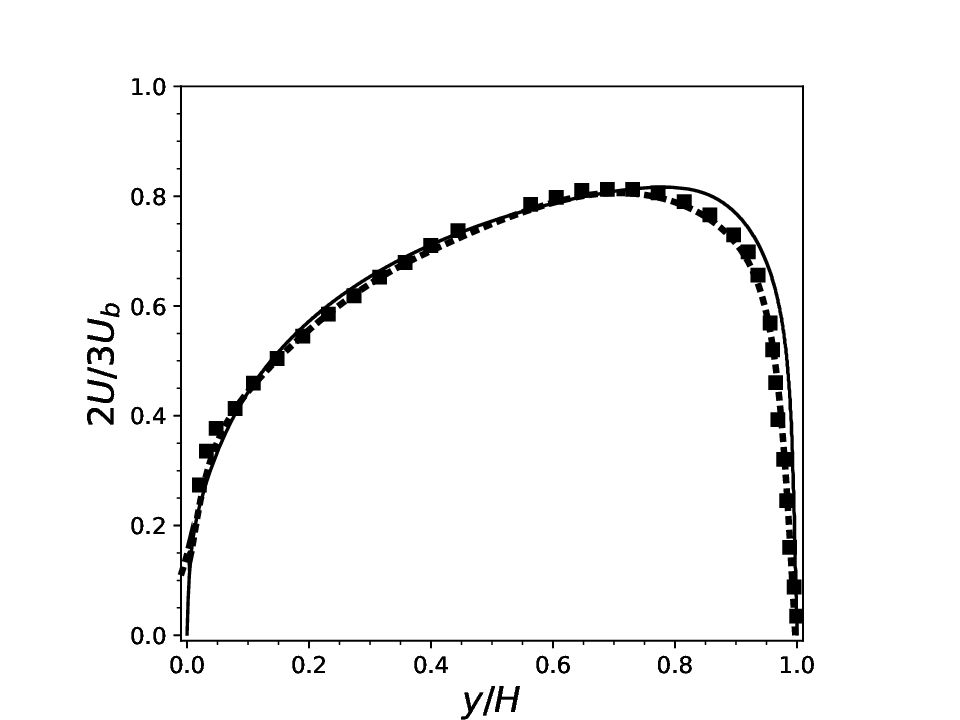}
\subcaption{$\omega=0$ boundary condition}
\end{subfigure}
\begin{subfigure}{0.48\textwidth}
\includegraphics[width=\textwidth]{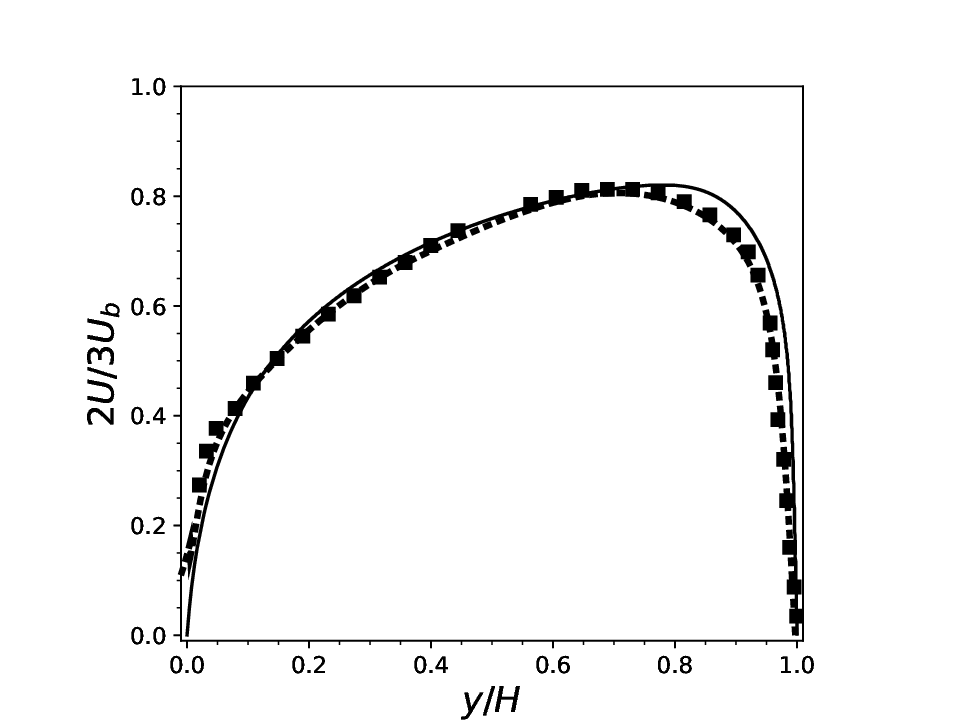}
\subcaption{Shear stress boundary condition}
\end{subfigure}
\begin{subfigure}{0.48\textwidth}
\includegraphics[width=\textwidth]{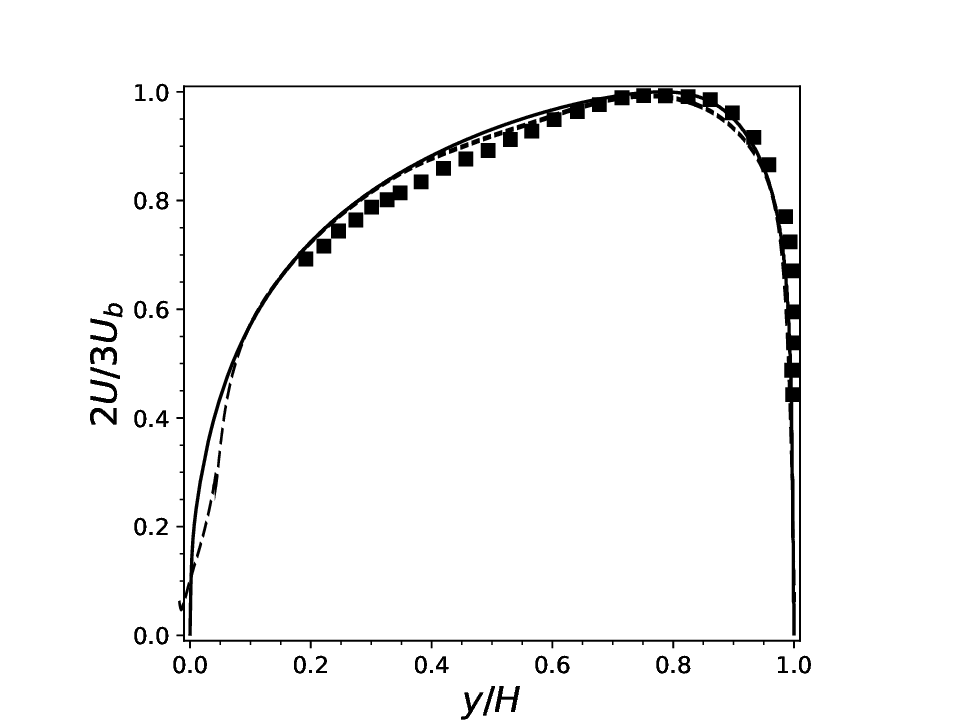}
\subcaption{$\omega=0$ boundary condition}
\end{subfigure}
\begin{subfigure}{0.48\textwidth}
\includegraphics[width=\textwidth]{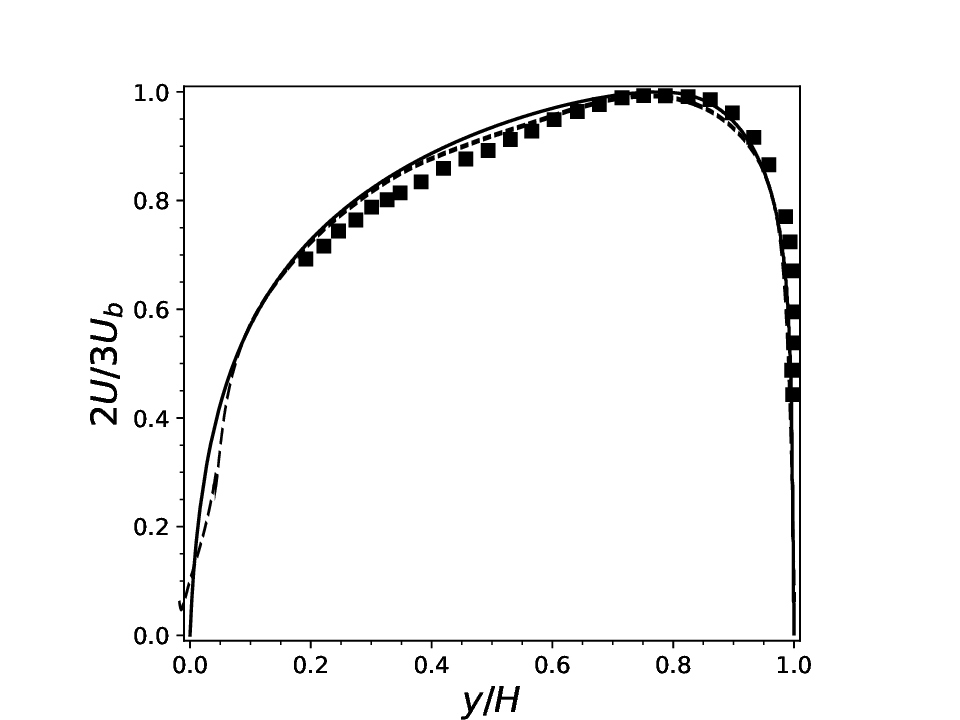}
\subcaption{Shear stress boundary condition}
\end{subfigure}
\caption{Rib-roughened channel.  $y=0$ is rough, $y=1$ is smooth. 
(a) and (b) dashed lines and symbols from \citet{Umair}, $R_\tau=1,100$, $r_+=1,100$;
(c) and (d) data from \citet{Ikeda},  $Re_{b}=18,500$, $r_+=1,000$.}
\label{fig:asymmetric}
\end{figure}

The mean velocity profile in a channel with one smooth and one rough wall is asymmetric.   Higher drag on the rough wall creates lower velocity on that side of the profile.
This configuration has been used to validate DNS \citep{Ikeda,Umair}. 
\Cref{fig:asymmetric} compares computations with the present model to data.   
The roughness is 2-D ribs, which create a sandgrain roughness that is an order of 
magnitude larger than the geometrical roughness.  In this case, sandgrain roughness is simply characterizing the log-law offset.  For 3-D roughness the geometrical and
sandgrain roughnesses are of similar magnitude.

Surface roughness can cause an adverse pressure gradient flow, that would be attached to a smooth wall, to separate from a rough wall \citep{Roughke}.
That can be explained as the consequence of roughness thickening the boundary layer: the pressure gradient
acceleration parameter is proportional to boundary layer thickness, which then increases, promoting separation. 
The experiment of \citet{Song2002} was intended to illustrate this:
\revised{ indeed, these experiments were conducted for the purpose of testing 
roughness models \citep{Roughke}.}
 First, measurements were made over a downward sloping,
smooth ramp. Then the ramp was covered with sandpaper and measurement were made over the rough wall.

\revised{The region of interest is a
flat plate followed by a contoured ramp. The unit of length is the ramp length, $L$,
The ramp height is $0.3L$.  
The flat plate lies between $-2L$ and 0, in the computational domain (the experiment
begins further upstream).
The ramp is a $33^{\circ}$, circular arc with a radius $1.81L$, beginning at 0 and
ending at 1. The upper wall is flat. The tunnel height expands from 
1.87 to $2.17L$.  
}  

\begin{figure}[ht]
\centering
\begin{subfigure}{0.5\textwidth}
\includegraphics[width=\textwidth]{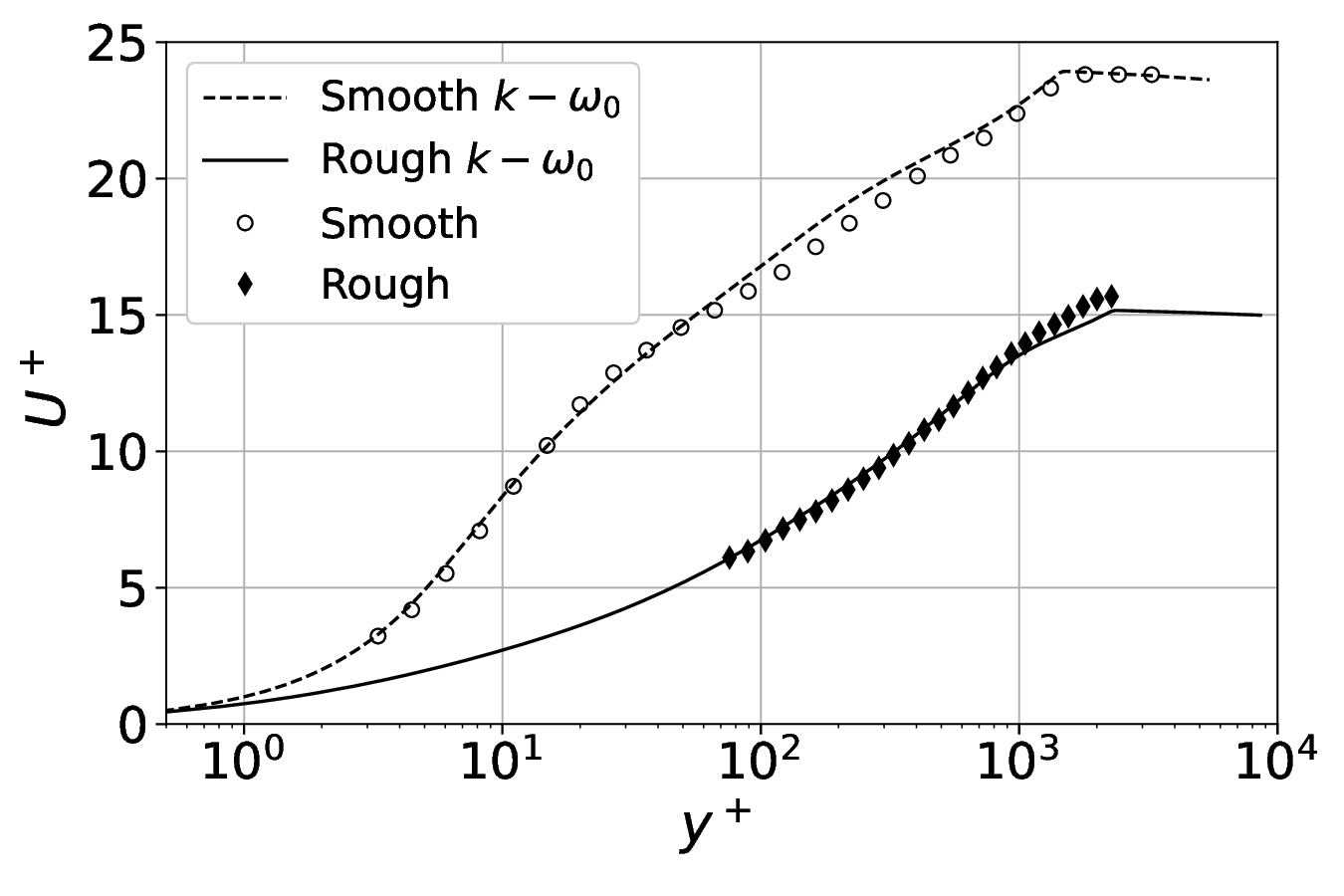}
\subcaption{\revised{Upstream profiles}}
\label{fig:songa}
\end{subfigure}\\
\begin{subfigure}{0.75\textwidth}
\includegraphics[width=\textwidth]{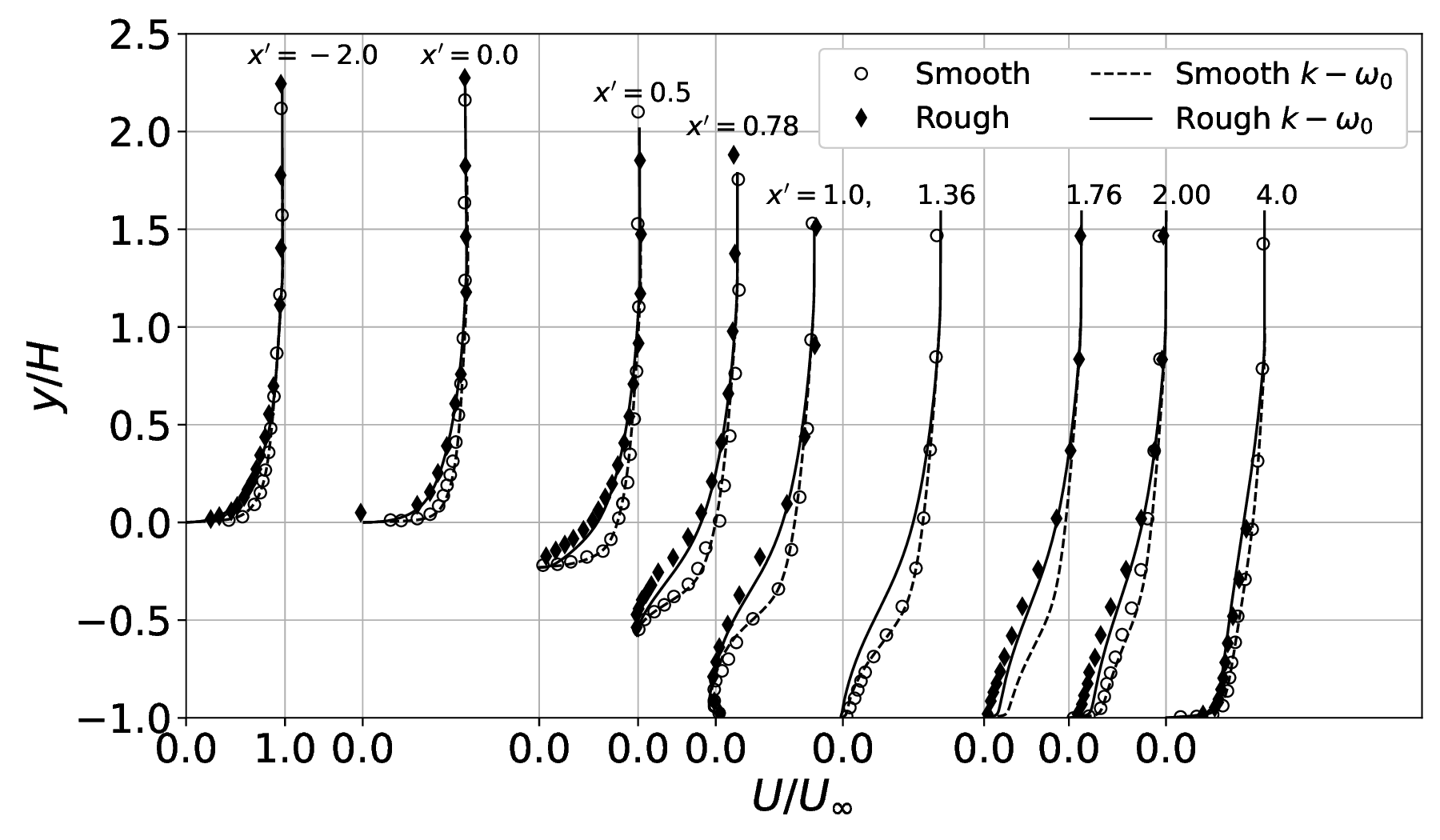}
\subcaption{\revised{Flow over ramp}}
\label{fig:songb}
\end{subfigure}
\caption{Velocity profiles for the smooth and rough ramp.}
\label{fig:song}
\end{figure}

\revised{In the numerical computation, a short region before the ramp is added, to generate 
a developed turbulent boundary layer, with slightly favorable pressure gradient.
The length of that region is adjusted to match the 
experimental momentum-thickness Reynolds numbers of 3,400 for the smooth case and 
3,900 for the rough case.
The $\omega_{0}=0$ boundary condition was used in this calculation. 
$\ell_{+}=5.8$ was found by matching the log-layer shift that is reported by \citet{Song2002}.  The 
corresponding $r_+=367$ can be read from \cref{fig:calib}. The velocity profile at $x/L = -2.0$ is shown 
in \Cref{fig:song}(a).}

\Cref{fig:song} (b) compares RANS computations to the measured profiles in rough and smooth cases.  
The smooth wall computation is in very good agreement with the experiment.  
The rough wall results show some inaccuracies, but the contrast between smooth and rough conditions is captured.   The tendency of roughness to cause separation is shown by the model.

\revised{It has been debated whether equivalent sandgrain roughness remains a valid representation
in strong disequilibrium \citep{JoTreview}.  This has not been resolved.  Here, the approach is
is to regard $r_{+}$ as an effective hydrodynamical property of the surface, irregardless of
flow conditions.}

\section{Discussion}

The idea of accommodating roughness by introducing an effective origin is attractive for its simplicity and 
has a physical motivation.  The irregular surface is represented by a smooth surface, at a location that depends
on the height of the surface asperities. 
This artificial, smooth surface could be regarded as a mean line through the actual, irregular surface.
Alternatively, it has been proposed in the literature that the effective surface should be
at the center of force on the asperities.  
The requirements of turbulence modeling allow this to be vague.  One could
describe the present approach as prescribing the artificial
 surface to be at height $\ell$, above the bottom of the asperities.  However, that geometrical
 interpretation is misleading.  The effective origin is used to prevent the eddy viscosity
 from becoming zero at the lower boundary, 
 because the roughness creates shear layers and turbulence near the surface.  $\ell$ is
 determined by turbulent mixing, not by geometry.
 
The $k-\omega_0$ model was developed as a means to avoid computing a singular function.  The  regular solution for $\omega_0$ is mapped onto a singular function, because the $1/y^2$ singularity
is an essential property of the \komeg model.  The $1/y^2$ singularity is derived from the balance between
dissipation and molecular diffusion.  That balance is upset 
by roughness; roughness introduces turbulence near the wall, overwhelming
molecular diffusion.  The present method of representing this 
is to alter the near-wall behavior of $\omega$ 
to $1/(y+\ell)^2$.  Also, $k$ is no longer 0 at the wall.  In combination, the model is able to extrapolate
from smooth to fully rough conditions.  
\revised{The roughness modification is developed for two possible $\omega_0$-boundary conditions. 
The $\omega_0(0) = 0$ variant offers a simpler implementation.}

In experiments, a virtual origin is needed to fit  measurements of $U$ to the log-law.
This is distinct from the effective origin, added to the turbulence model.  The virtual origin is determined empirically.
In the present formulation, it is possible to derive a formula for the virtual origin,
 under fully rough conditions.

\bibliography{reference}

\end{document}